\documentclass[a4paper,12pt]{article}
\usepackage[utf8]{inputenc} 
\usepackage[dvips]{color,graphicx} 
\usepackage{mathrsfs} 
\usepackage{amssymb} 
\usepackage{amsmath} 
\usepackage{feynmp} 
\usepackage{slashed} 
\usepackage{latexsym} 
\usepackage{graphicx} 
\usepackage{verbatim} 
\usepackage[normalem]{ulem} 
 
\newcommand{\be}{\begin{equation}} 
\newcommand{\ee}{\end{equation}} 
\newcommand{\ba}{\begin{array}} 
\newcommand{\ea}{\end{array}} 
\newcommand{\bea}{\begin{eqnarray}} 
\newcommand{\eea}{\end{eqnarray}} 
\newcommand{\balg}{\begin{align}} 
\newcommand{\ealg}{\end{align}} 
\newcommand{\bit}{\begin{itemize}} 
\newcommand{\eit}{\end{itemize}} 
\newcommand{\trm}[1]{\textrm{#1}} 
\newcommand{\mbf}[1]{\mathbf{#1}} 
\newcommand{\tbf}[1]{\textbf{#1}} 
\newcommand{\mcl}[1]{\mathcal{#1}} 
\newcommand{\mbb}[1]{\mathbb{#1}} 
\newcommand{\msc}[1]{\mathscr{#1}}

\newcommand{\Mpc}{\trm{\Mpc}} 
\newcommand{\yr}{\trm{\yr}} 
\newcommand{\eV}{\trm{\eV}}

\newcommand{\Diag}[1]{\trm{diag}\{ #1 \}}

\newcommand{\vth}{\vspace{.3cm}}

\newcommand{\tr}[1]{\textrm{Tr}[#1]}

\begin{document} 
 
\title{
\Large \bf  
Relating neutrino masses and mixings by discrete symmetries}  
\author{ 
{D. Hernandez \thanks{email: 
\tt dhernand@ictp.it}~~\,and 
\vspace*{0.15cm} ~A. Yu. Smirnov \thanks{email: 
\tt smirnov@ictp.it} 
} \\ 
{\normalsize\em The Abdus Salam International Centre for Theoretical     
Physics} \\ 
{\normalsize\em Strada Costiera 11, I-34014 Trieste, Italy  
\vspace*{0.15cm}} 
} 
\date{} 
\maketitle 
\thispagestyle{empty} 
\vspace{-0.8cm} 
\begin{abstract} 
 
Lepton mixing can originate from the breaking of a flavor symmetry  
in different ways in the neutrino and the charged lepton sector. 
We propose an extension of this framework which allows to connect the mixing parameters  
with masses, and more precisely, with certain types of degeneracy of the neutrino mass spectrum.   
We obtain relations between the mixing parameters for the cases  
of  partial degeneracy, $m_1 = m_2$,  and  complete degeneracy, $m_1 = m_2 = m_3$.  
These relations determine also the  Majorana phases.  
It is  shown that relatively small corrections to these lowest order results   
can produce the required mass splitting and modify the mixing without significant changes of the
other symmetry  results.

\end{abstract}

\vspace{1.cm} 
\vspace{.3cm} 

\newpage 
 
 
\section{Introduction} \label{sec1} 
 
 
Lepton mixing can be a consequence of the breaking of a   
flavor symmetry $G_f$ into different \emph{residual symmetries}, 
$G_\nu$ and $G_\ell$, for the neutrino and charged lepton mass matrices  
respectively~\cite{altarelli-feruglio}.   
To a large extent,  this approach was motivated by the peculiar values of the lepton 
mixing parameters that seemed to be well  approximated by the so called tri-bimaximal (TBM)  
mixing~\cite{tbm}.   
TBM turned out to be difficult to connect with  the known ratios of masses of the charged  leptons and  
neutrinos, and therefore, in this approach, the masses 
of neutrinos and charged leptons ``decouple''   from mixing. In other words,  TBM implies the form invariance  
of the mass matrix - a situation in which mixing is determined by symmetry alone.  Models that reproduced  
the TBM and other interesting mixing patterns were  built along these  
lines~\cite{A4TBMmodels, S4BMmodels,S4TBMmodels,A5GRmodels,toorop}.  
Independent physics (additional symmetries) was assumed to be responsible  
for the  hierarchy of masses of leptons. Indirect relations between mixing and mass spectrum  
appear in some specific models, as a result of model structure and particle content. 
 
The separate description of mixing and mass   hierarchies (ratios of masses) with different physics involved  
looks unsatisfactory.  
Indeed, 
 
\begin{enumerate} 
 
\item In general, masses and mixing have the same  origins following formally  
from diagonalization of the mass  
matrices. They are generated by the same type of Yukawa couplings and so should be somehow related. 
 
\item In the quark sector various relations between the mass rations and mixing 
parameters have been uncovered with the Gatto-Sartori-Tonin relation \cite{GST} 
being the most appealing one. The latter can be explained by 
an abelian flavor symmetry, e.g., in the Froggatt-Nielsen approach \cite{FN}. 
There were some attempts to use discrete symmeries to produce   
relations between masses and mixing (see e.g. \cite{Branco}). 
 In a number of recent models with discrete flavor symmetries the  
sum rules for neutrino masses  
(linear relations between masses or their  inverse quantities)  
have been realized (see e.g. \cite{kingg}). However, these relations do not depend on mixing  
and  turn out to be consequences  
of specific restricted model contents and vacuum alignments and do not  
follow from residual symmetries of the neutrino mass matrix.

\item The now established relatively large 1-3 mixing and the indications 
of significant deviations of the 2-3 mixing from maximal 
\cite{globalfit} rule out the exact TBM, and therefore cast doubts 
to the explanation of mixing separated from masses  
through nonabelian symmetries. At the same time, it was shown in   
\cite{ge, first-paper, second-paper, lam} that  flavor symmetries 
can still accommodate the recent results on the 1-3 and 2-3 mixings. 
\end{enumerate} 
 
In ~\cite{first-paper, second-paper} we proposed a formalism  for 
``symmetry building''  which relies 
on the aforementioned partial breaking of $G_f$ into two 
subgroups, $G_\ell$ and $G_\nu$.  We 
used it to  obtain relations between the mixing parameters without explicit 
reference to any particular model.  
It was shown that consequences of symmetries for mixing, that is,  
relations between the mixing 
parameters or elements of mixing matrix (at least at the lowest order)   
can be obtained immediately once residual symmetries (transformations) of the neutrino 
and charged lepton mass matrices are known.  
These symmetry group relations can be obtained without model building and  
explicit construction of the mass matrices and their diagonalizations. 
Essentially, it is only assumed that the model 
is constructed and it leads to mass matrices with given symmetry properties. 
The relations in  \cite{first-paper, second-paper}   
can be viewed as a tool of symmetry or/and model building. 
Once the required relations  and the corresponding residual 
symmetries are identified, one can come back and construct the corresponding 
complete flavor symmetry and model. After a model is constructed one can compute corrections 
to zero order structures. In general, this formalism does not allow to  
compute the latter model-dependent corrections.
 
In this paper we further develop  
this formalism, which in ~\cite{first-paper, second-paper} was elaborated  
for mixing only, in order to include also neutrino masses. Consequently,  
we obtain relations between the mixing parameters and certain types 
of the neutrino mass spectrum. 
 
The crucial point of the explanation of mixing decoupled from masses was 
to use, as $G_\nu$, the symmetry of a generic neutrino mass matrix $M_\nu$ 
with arbitrary eigenvalues. For Majorana neutrinos the maximal  
symmetry of $M_\nu$  is ${\bf Z}_2 \times {\bf Z}_2$, as can be seen immediately in the 
basis where the mass matrix is diagonal\footnote{A complete scan of the groups  
with order less than 1536 and $G_\nu = Z_2 \times Z_2$ was performed in \cite{lindner}.}.  
This generic symmetry does not constrain the masses  and 
therefore leads to the decoupling of mixing.  Hence, in  
order for the symmetry to predict both masses and mixings,   
$G_\nu$ should be enlarged in such a way that invariance  
of the mass matrix  is satisfied only for certain mass  
spectra.  
 
In this paper, we focus on residual symmetries $G_\nu$ that  
lead to equalities of the neutrino masses.   
 Indeed, unitary symmetry transformations  
can lead either to equalities of the masses or to zero values  
of masses\footnote{The possibility of the symmetry leading 
to vanishing masses has been considered in \cite{patel}.}. 
In this connection we will explore two possiblities: (i) two degenerate neutrinos, 
{\it i.e.}, equality of two masses, (ii) three degenerate neutrinos.  
The first case can be considered as the lowest order  
approximation  to spectra with  both normal and inverted mass hierarchies.  
Then, corrections are required which 
lead to  splitting between the masses.  
In the case of normal mass hierarchy with  
two vanishing masses, the corrections should generate the  mass   
of at least the second neutrino.

The paper is organized as follows.  
In Sec.~\ref{sec2} we describe the model  independent method for ``symmetry building'',  
applied here to the case of specific  neutrino mass spectrum.  
In Sec.~\ref{sec3}  we consider the case of partial degeneracy 
(equality of two masses). We derive the relations between 
mixing parameters which  also  include the Majorana phases. 
In Sec.~\ref{sec4} we derive constrains on the mixing and phases 
in the case of completely degenerate spectrum. Discussion and conclusions 
are presented in Sec.~\ref{sec5}.

 
\section{Symmetry relations for mixing  and masses} 
\label{sec2} 
 

We assume that neutrinos are Majorana particles.   
Working in the flavor basis, the mass terms of the lepton sector of the  
Lagrangian read 
\be 
\msc{L}_{mass} = \bar{E}_R m_\ell \ell_L + \frac{1}{2}   
\bar{\nu^c}_{f L} M_{\nu U} \nu_{f L}    
\label{lag} + \trm{h.c.} \, , 
\ee 
where  $\nu_{f L}$, $\ell_L$ and $E_R$ are the leptonic fields:  
$\nu_{f L} \equiv (\nu_e,\, \nu_\mu,\,\nu_\tau )_L^T$,  
$\ell_L \equiv (e,\,\mu,\,\tau)_L^T$,  $E_R \equiv (e,\,\mu,\,\tau)_R^T$,   
and  $m_\ell \equiv \trm{diag}\{m_e,\,m_\mu,\,m_\tau\}$.  
The flavor neutrino states are related to the mass eigenstates,   
$\nu_L \equiv (\nu_1,\,\nu_2,\,\nu_3)_L^T$, by the  
Pontecorvo-Maki-Nakagawa-Sakata mixing matrix:  $\nu_{f L} = U_{PMNS} \nu_L$.   
Then, the neutrino mass matrix in the flavor basis, $M_{\nu U}$,  
can be expressed via the diagonal mass  matrix  
of the neutrino mass eigenvalues, $m_{\nu } \equiv \trm{diag}\{m_1,\,m_2,\,m_3\}$,  
and $U_{PMNS}$ as  
\be 
M_{\nu U} = U_{PMNS}^* m_{\nu } U^\dagger_{PMNS}\, .  
\label{numass} 
\ee 
 
Let $\mcl{G}_\ell$ and $\mcl{G}_{\nu U}$ be the groups of symmetry transformations  
that leave invariant the charged lepton and neutrino  
mass terms in Eq.~\eqref{lag}. The residual flavor symmetries  
in the charged lepton and neutrino sectors are in general finite subgroups  
of these: $G_\ell \subset \mcl{G}_\ell$ and  
$G_\nu \subset \mcl{G}_{\nu U}$. We proceed to identify $G_\nu$ and $G_\ell$ 
systematically. 
 
For the charged leptons we have $\mcl{G}_\ell \equiv U(1)^3$  
corresponding to the electron, muon and tau lepton numbers.  
As in our previous papers \cite{first-paper, second-paper},    
we assume that $G_\ell = \mbf{Z}_m$.  
The fact that $\mbf{Z}_m$ is generated by one element implies that it leads to a minimal number of constraints on the mixing matrix. 
Analysis of bigger groups, that lead to stricter conditions $U_{PMNS}$, is beyond the scope of this paper. 
 
A representation of $G_\ell$ is given by the matrix $T$  
such that $\ell_L$ and $E_R$ transform as  
\be 
\ell \rightarrow T \ell_L \,,\quad E_R \rightarrow T E_R\,,  
\label{cl-transf} 
\ee 
where 
\be 
T \equiv \Diag{e^{i\phi_e},\, e^{i\phi_\mu},\, e^{i\phi_\tau}} \label{Tdef}  
\ee 
and 
\be 
\phi_\alpha \equiv 2\pi \frac{k_\alpha}{m} \,, \quad   \alpha = e,\, \mu,\, \tau \;. \label{phidef} 
\ee 
Invariance of the charged lepton mass matrix $m_\ell$ under $T$ means that the following equality holds:  $T m_{\ell} T^{\dagger} = m_{\ell}$.  
According to Eqs.~\eqref{Tdef} and \eqref{phidef}, $T$ satisfies the condition $T^m = \mbb{I}$.  
It is enough  to consider a  subgroup of $SU(3)$ as the flavor group. This restriction simplifies the considerations without having any impact in the results of the paper. One can show that the additional $U(1)$ of $U(3)$ can be factored out and does not induce any constraint on mixing.  
Thus, we  impose the equality  
$$ 
\phi_e + \phi_\mu + \phi_\tau = 0,  
$$ 
or equivalently, $k_\tau = - k_e + k_\mu$, that ensures $\trm{Det}[T] = 1$. 
 
\vth 
 
Considering now the symmetry group $\mcl{G}_\nu$  
of the Majorana mass term of neutrinos,   
we explore the possibility of approximate degeneracy of the mass spectrum.  
In the neutrino mass basis the invariance of mass matrix  
under the transformation 
\be 
\nu_{L} \rightarrow S \nu_L \, ,  
\label{nu-sim} 
\ee  
where $S$ belongs to the group $\mcl{G}_\nu$, means that   
\be 
S^T m_{\nu } S = m_\nu \,  
\label{m-inv-cond}.  
\ee 
If $S$ satisfies Eq.~\eqref{m-inv-cond}, the corresponding 
matrix $S_U$ that leaves $M_{\nu U}$ invariant, {\it i.e.} 
$$ 
S_U^T M_{\nu U} S_U = M_{\nu U} \,, 
$$ 
can be found by switching to the flavor basis:  
\be 
S_U = U_{PMNS}SU_{PMNS}^\dagger \,. \label{S-flavorbasis} 
\ee 
Hence, the group $\mcl{G}_{\nu U}$ is obtained from $\mcl{G}_\nu$ by applying a  
similarity transformation with $U_{PMNS}$ to all elements of $\mcl{G}_\nu$.  
The residual symmetry $G_\nu$ is a discrete subgroup 
of $\mcl{G}_{\nu U}$.

\vth 
 
Three cases can be distinguished for $\mcl{G}_\nu$  
with increasing symmetry that corresponds to increasing degree  
of degeneracy of $m_\nu$.  
 
\begin{itemize} 
\item[\tbf{A}.] \emph{No degeneracy}: $\mcl{G_\nu} = \mbf{Z}_2\otimes\mbf{Z}_2$. 
 
The diagonal neutrino mass matrix $m_{\nu}$ with arbitrary  
eigenvalues is invariant under the transformations (\ref{nu-sim})   
with   
\be 
S_1 = \Diag{1,\,-1,\,-1}\,,\quad S_2 = \Diag{-1,\,1,\,-1}\,,\quad  
\label{Si} 
\ee 
and  $S_3 = S_1 S_2$.  
This case was analyzed in \cite{first-paper, second-paper}  
and $G_\nu = \mbf{Z}_2$ or $G_\nu = \mbf{Z}_2\otimes \mbf{Z}_2$ 
are possible.

\item[\tbf{B}.] \emph{2 degenerate neutrinos}:  
$\mcl{G}_\nu = SO(2) \otimes \mbf{Z}_2$. 
 
In addition to the matrices in Eq.~\eqref{Si},  
the neutrino mass matrix is invariant under rotations  
of the plane of mass degeneracy\footnote{A model for two degenerate neutrinos based on the group $D_N$ - a discrete subgroup of $O(2)$ - has been proposed in~\cite{Araki:2012hb} where a relation which connects  $\theta_{13}$, the electron mass and the CP-phases was obtained. The approach in~\cite{Araki:2012hb} differs from ours in that $D_N$ is the symmetry of whole leptonic sector, introduced to explain small observed parameters which describe masses and mixing.}. Taking into account the measured  
neutrino mass differences, the approximate degeneracy  
must be between $m_1$ and $m_2$. Therefore  we take 
\be 
m_{\nu 3} \simeq \left( \begin{array}{ccc} 
m && \\ 
& m & \\ 
&& m' \end{array} \right),  
\label{mnuB} 
\ee 
which is invariant under the transformation   
\be  
S \equiv S_{\zeta}  = \left( \begin{array}{ccc} 
c_\zeta & -s_\zeta & \\ 
s_\zeta & c_\zeta & \\ 
& & 1 
\end{array} \right) \label{SB} \, ,  
\ee 
where $c_\zeta \equiv \cos \zeta$, $s_\zeta \equiv - \sin \zeta$. 
Analogously to the case of charged leptons, 
we impose $G_\nu = \mbf{Z}_n$ so that $S_\zeta^n = \mbb{I}$ 
and  $\zeta = 2\pi l/n$. In the flavor basis, Eq.~\eqref{S-flavorbasis}, we must also have 
$$ 
S_U^n = \mbb{I} \,. \label{S^m} 
$$

\item[\tbf{C.}] \emph{3 degenerate neutrinos}:  
$\mcl{G_\nu} = SO(3)$. 
 
In this case we have 
\be 
m_\nu \simeq m \mbb{I} ,   
\label{mnuC} 
\ee 
and $S$ can be any orthogonal $3\times 3$ matrix.  
We note that, according to the Euler rotation theorem,  
any 3D rotation is a 1D rotation around a certain axis.  
Thus, any matrix $S$ that is a symmetry of $m_\nu$ can be written as 
\be 
S \equiv S_{\zeta O} = OS_\zeta O^T , 
\label{SzetaO} 
\ee 
where $O$ is an orthogonal matrix.   
The Euler rotation theorem implies that   
a basis in the neutrino sector can be always selected such  
that the matrix $S$ in this basis has the form $S_\zeta$.  
Turning the argument around,  
imposing only a $\mbf{Z}_n$ symmetry generated by $S_{\zeta O}$  
is not enough to force the three neutrinos to be degenerate.  
The most general mass matrix that is left invariant  
by a $\mbf{Z}_n$ subgroup of $O(3)$ has only two equal    
eigenvalues. Thus, if $G_\nu$ imposes full degeneracy of  
the neutrino mass matrix, it must be one of the non-Abelian  
subgroups of $O(3)$ with 3D representations, {\it i.e.},  
$\mbf{A}_4$, $\mbf{S}_4$ or $\mbf{A}_5$\footnote{For an early $\mbf{A}_4$ 
model predicting a nearly degenerate neutrino spectrum, see \cite{ma-degenerate} }.

\end{itemize}

\vth

As it was shown in~\cite{first-paper},  the relations between mixing matrix elements  
follow from the condition  
that the symmetry transformations of the charged leptons 
and neutrinos in the flavor basis  belong to the same discrete group $G_f$. That is, the product  
\be   
W_{U} \equiv  S_{ U}T \,   
\label{rel2} 
\ee 
must also belong to $G_f$. Furthermore, since $G_f$ is {\it finite},  
there must exist an integer $p$ such that  
\be 
W^p_{U} = (S_{U}T)^p = \mbb{I}.  
\label{wrelation} 
\ee 
The  relations  
\be 
S_{U}^n = T^m = W^p_{U} = \mbb{I}    
\label{rel1} 
\ee 
form a presentation of $G_f$ and define the von Dyck group $D(n,m,p)$.  
 
Eq.~\eqref{wrelation} is a constraint on the mixing matrix~\cite{first-paper}. To see  
this, notice that the eigenvalues of $W_U$ are three $p$-th  
roots of unity, $\lambda_1^{(p)}$, $\lambda_2^{(p)}$ and $\lambda_3^{(p)}$. Defining 
$$ 
a \equiv \tr{W_U} = \lambda_1^{(p)} + \lambda_2^{(p)} + \lambda_3^{(p)} 
$$ 
we have from Eq.~\eqref{wrelation} that 
\be 
\tr{U_{PMNS}S U_{PMNS}^\dagger T} \equiv \tr{W} = a \,. \label{witha} 
\ee 
Since the $p$-th roots of unity are a finite set,  
the RHS of this equation takes values from  
a finite set of known complex numbers.  
For known $a$, and given $S$ and $T$, Eq.~\eqref{witha}  
is a complex condition that the entries of $U_{PMNS}$ must satisfy.

Although we will proceed below in all generality,  
the case analysis of $n$, $m$ and $p$ is significantly  
reduced after the following consideration. It is a known fact  
\cite{first-paper} that in order for the von Dyck group  
to be finite, one of $n$, $m$ or $p$ must be equal to 2.  
In \cite{first-paper, second-paper}  
we took $n=2$ consistent with $G_\nu = \mbf{Z}_2$.  
However, in order to enforce degeneracy in the neutrino  
mass matrix, it must be $n \geq 3$. Assuming also  
that  all charged leptons have different charges under $T$, \emph{i. e.} 
 $m \geq 3$, we obtain that due to the finiteness of the group  
it must be necessarily $p=2$. 
 
Eq.~\eqref{wrelation} can then be written as 
$$ 
W_U^2 = (U_{PMNS} S U_{PMNS}^\dagger T)^2 = \mbb{I},  
$$ 
and the eigenvalues of $W_U$ must be equal to $+1$ or $-1$.  
Moreover, taking into account that $\det[W_U] = 1$,  
the eigenvalues of $W_U$ must be $\{1,\,-1,\,-1\}$,  
if $W_U$ is not trivial. Hence, we obtain 
\be 
a = \tr{W_U} = \tr{U_{PMNS} S U_{PMNS}^\dagger T} = -1 \,.  
\label{main} 
\ee 
The condition in Eq.~\eqref{main} is appropriate when  
a residual symmetry in the neutrino sector forces the neutrino   
mass matrix to be of the form B or C.

\vth

In what follows  
we find explicitly the constraints imposed on $U_{PMNS}$  
and compare them with experimental data. For $U_{PMNS}$ we will use the standard  
parametrization given by 
\begin{eqnarray} 
U_{PMNS}& = & U_{23} (\theta_{23})\Gamma_\delta U_{13}(\theta_{13}) \Gamma_\delta^* U_{12}(\theta_{12})  
\Gamma_M  = \\ 
& = &  
\left( \begin{array}{ccc} 
c_{12}c_{13} & s_{12}c_{13} & e^{-i\delta}s_{13}  \\ 
-s_{12}c_{23}-e^{i\delta}c_{12}s_{23}s_{13} & c_{12}c_{23}  
- e^{i\delta}s_{12}s_{23}s_{13} & s_{23}c_{13} \\ 
s_{12}s_{23}-e^{i\delta}c_{12}c_{23}s_{13} &  
-c_{12}s_{23}-e^{i\delta}s_{12}c_{23}s_{13} & c_{23}c_{13}  
\end{array}\right) \,\Gamma_M ,   
\label{pmnspar} 
\end{eqnarray} 
where  $U_{ij}$ are the matrices of rotations in the $ij-$planes on the  
angles $\theta_{ij}$,   
\be 
\Gamma_\delta \equiv \trm{diag} \{1,\, 1, \,e^{i\delta}\} ,  ~~~~~  
\Gamma_M \equiv \trm{diag} \{1,\,e^{i\kappa},\,e^{i\chi}\} \, ,  
\ee 
and $c_{12} \equiv \cos\theta_{12}$, $s_{12} \equiv \sin\theta_{12}$, {\it etc.}.

\section{Constraints on mixing for the partially degenerate spectrum} 
\label{sec3} 

For partially degenerate spectrum, the neutrino  
mass matrix and the corresponding symmetry  
are given by Eqs.~\eqref{mnuB} and \eqref{SB} respectively,  
with $\zeta = 2\pi l/n$.  The matrix $m_\nu$ can be a good lowest order  
approximation to both normal  and inverted  mass hierarchies. Corrections could then  
produce small splitting between the degenerate states and modify mixing angles when  
needed.   
 
Setting $S = S_\zeta $ we have from (\ref{main}) the symmetry relation 
$$
\tr{S_U T} =  \tr{U_{PMNS}S_\zeta U^\dagger_{PMNS}T} = - 1.  
$$ 
Explicit computation of  $S_U$   
(see Eq.~\eqref{S-flavorbasis}) gives 
\be 
(S_{U})_{\alpha \alpha} =  c_\zeta +  2 s_{\zeta/2}^2 |U_{\alpha 3}|^2 +  
i  2 s_\zeta \trm{Im} (U_{\alpha 2}U_{\alpha 1}^*),  
\label{eq:salpha1} 
\ee 
where $s_{\zeta/2} \equiv \sin (\zeta/2)$.   
It is convenient to introduce the real and imaginary parts of    
$(S_{U})_{\alpha \alpha} = R_\alpha + i I_\alpha$:   
\be 
R_\alpha = c_\zeta +  2 s_{\zeta/2}^2 |U_{\alpha 3}|^2 \,,\quad  
I_\alpha = 2 s_\zeta {\rm Im} (U_{\alpha 2}U_{\alpha 1}^*).  
\label{randi} 
\ee 
Notice that the index $3$ in the real part  
of $(S_{U})_{\alpha \alpha}$ is related  
to the non-degenerate third mass eigenstate. According to Eq.~\eqref{witha}, the trace of $W_U$ equals  
$$ 
{\rm Tr}(W_{U}) =  (S_{U})_{ee}e^{i\phi_e} +  
(S_{U})_{\mu \mu} e^{i \phi_\mu} + (S_{U})_{\tau \tau} e^{i\phi_\tau} =  a,  
$$ 
and consequently, from the real and imaginary parts of this equation   
we obtain using Eq.~(\ref{randi})   
\be  
\sum_\alpha \left(R_\alpha\cos\phi_\alpha -  
I_\alpha\sin\phi_\alpha \right) = \trm{Re}[a] \,,\quad \alpha = e,\,\mu,\,\tau \,, 
\label{realp} 
\ee 
\be   
\sum_\alpha \left( R_\alpha\sin \phi_\alpha -  
I_\alpha\cos\phi_\alpha\right)   
= \trm{Im}[a] \,,\quad \alpha = e,\,\mu,\,\tau \,. 
\label{imp} 
\ee 
These are the conditions imposed on mixing by the symmetry in the case of partial  
degeneracy.  
Explicit equations for the mixing angles and phases  
in Eq.~\eqref{pmnspar} can be found by substituting  
$R_\alpha$ and $I_\alpha$ from Eq.~\eqref{randi} in Eqs.~\eqref{realp} and \eqref{imp}. 
 
As an example, we consider the case in which  
$\phi_e = 0$, $\phi_\mu = - \phi_\tau \equiv \psi$, so that  
the charged lepton transformation matrix has the form  
\be 
T  = {\rm diag} \left \{1 , e^{i \psi}, e^{- i \psi} \right\}.    
\label{Talpha} 
\ee 
Eqs.~\eqref{realp} and \eqref{imp} are then reduced to 
\be 
R_e + c_\psi (R_\mu + R_\tau) + s_\psi (I_\tau - I_\mu) = \trm{Re}[a] \,,   
\label{rp} 
\ee 
\be   
I_e + c_\psi (I_\mu + I_\tau) + s_\psi (R_\mu - R_\tau) = \trm{Im}[a]\,,   
\label{ip} 
\ee 
and inserting Eq.~\eqref{randi} in \eqref{rp} and  \eqref{ip}  we obtain  
\be 
|U_{e 3}|^2  - 2 x\, 
{\rm Im} \left[U_{\mu 2} U_{\mu 1}^* - U_{\tau 2} U_{\tau 1}^* \right] +  
x^2 = \frac{1 + \trm{Re}[a]}{4s^2_{\psi/2}s^2_{\zeta/2}} \, ,   
\label{eq:reex} 
\ee 
\be 
2 \trm{Im}\left[U_{e 2} U_{e 1}^*\right] -  
y \left(|U_{\tau 3}|^2  - |U_{\mu 3}|^2 \right) = \frac{\trm{Im}[a]}{s_\zeta} \,.   
\label{eq:inex} 
\ee 
Here we have introduced parameters  
\be 
x \equiv \cot \frac{\psi}{2} \cot \frac{\zeta}{2} \,,\quad  
y \equiv \cot \frac{\psi}{2}  
\tan \frac{\zeta}{2}\, ,  
\label{eq:xy} 
\ee 
which depend only on the group properties.  
From Eq. (\ref{eq:xy}),  
$x y = \cot^2 \frac{\psi}{2}$ and $x/y = \cot^2 \frac{\zeta}{2}$,   
and consequently, $x$ and $y$ should have the same sign.   
 
Eqs.~(\ref{eq:reex}, \ref{eq:inex}) can be immediately  
generalized to the cases in which the lepton that  
has zero charge under $T$ is the muon or the tau. The general equations are 
\be 
|U_{\alpha 3}|^2  - 2 x \,{\rm Im} \left[U_{\beta 1} U_{\beta 2}^*  
- U_{\gamma 1} U_{\gamma 2}^* \right] + x^2 =   
\frac{1 + \trm{Re}[a]}{4s^2_{\psi/2}s^2_{\zeta/2}} ,  
\label{eq:reexg} 
\ee 
\be 
2 {\rm Im}\left[U_{\alpha 1} U_{\alpha 2}^*\right]  
- y  \left(|U_{\gamma  3}|^2  - |U_{\beta 3}|^2 \right)  
=  \frac{\trm{Im}[a]}{s_\zeta} \,. 
\label{eq:inexg} 
\ee 
Here, $(\alpha,\, \beta,\,\gamma)$ is  
$(e,\,\mu,\,\tau)$ or any other combination  with  
a cyclic permutation of these flavor indices.    
Eqs.~(\ref{eq:reex}), (\ref{eq:inex})  
correspond to the case $\alpha = e$. Notice that Eqs.~\eqref{eq:reexg}  
and \eqref{eq:inexg} represent yet another generalization  
of the results of \cite{first-paper} which can be reproduced by setting $\zeta = \pi$. 
 
When there is partial degeneracy in the neutrino mass matrix,  
only $a = -1$ (see Eq.~\eqref{main}) leads to finite groups.  
The explicit conditions on mixing imposed by the symmetry are then reduced to 
\bea 
(|U_{\alpha 3}| \mp x)^2 & = &  2 x\,  
\left({\rm Im} \left[U_{\beta 1} U_{\beta 2}^* -  
U_{\gamma 1} U_{\gamma 2}^* \right] \mp |U_{\alpha 3}|\right) \, , 
\label{eq:mainre}\\ 
2 {\rm Im}\left[U_{\alpha 1} U_{\alpha 2}^*\right] & = &   
y  \left(|U_{\gamma  3}|^2  -  
|U_{\beta 3}|^2 \right) \,. 
\label{eq:mainim} 
\eea 
 
The set of solutions of Eqs.~(\ref{eq:mainre}, \ref{eq:mainim})  
is very restricted. In order to show this, we will use    
the standard parametrization Eq.~\eqref{pmnspar}   
for  $|U_{\alpha i}|^2$ and consider for definiteness the case $\alpha = e$.  
Notice nonetheless that our results do not lose generality  
since for any choice of $\alpha$ there exists a parametrization  
such that the equations have the form to be discussed below. 
 
We distinguish two cases: $x > 0$ and $x < 0$   
which  imply $ y > 0$ and $y < 0$ respectively.  
In the standard parametrization and for  
$|U_{e 3}| =  \sin\theta_{13} \neq  0$, Eqs.~(\ref{eq:mainre}, \ref{eq:mainim})   
can be written as 
\begin{align} 
\left( \sin \theta_{13} \mp x \right)^2 & =  2 x\, (A_1 + A_2),   
\label{SPeq1} \\ 
\sin2\theta_{12}\sin\kappa & = y \cos2\theta_{23} .    
\label{SPeq2} 
\end{align} 
Here $\kappa$ is the Majorana phase defined 
in Eq.~\eqref{pmnspar} and the minus (plus) sign corresponds to  $x > 0$ $(< 0)$. $A_1$ and $A_2$ are given  
by 
\begin{align} 
A_1 & =  - \frac{1}{2}( 1 + \sin^2 \theta_{13})   
\cos2\theta_{23}\sin2\theta_{12}\sin\kappa ,  
\label{A1}\\ 
A_2 & = \sin\theta_{13}\left[ (\cos\kappa \sin\delta -  
\cos\delta\sin\kappa \cos2\theta_{12})\sin2\theta_{23} \mp 1 \right].  
\label{A2} 
\end{align} 
 
Substituting $\cos2\theta_{23}$ from Eq.~\eqref{SPeq2} into Eq.~\eqref{A1} we obtain 
\be 
A_1 = - \frac{1}{2y}(1 + \sin^2 \theta_{13})\sin^22\theta_{12}\sin^2\kappa \, , 
\label{aa1} 
\ee 
so that $A_1 \leq 0$ for $x > 0$ and $A_1 \geq 0$  for $x < 0$   
(recall that  $x$ and $y$ have the same sign).  Since  
$$ 
|\cos\kappa \sin\delta - \cos\delta\sin\kappa \cos2\theta_{12}| \leq 1 ,  
$$ 
from Eq.~\eqref{A2} we have  $A_2 \leq 0$ for $x > 0$ and $A_2 \geq 0$ for $x < 0$.  
We can combine  these two results:  
\bea  
A_i &  \leq & 0, ~~~  x > 0,  
\nonumber\\ 
A_i & \geq  & 0, ~~~ x < 0,    ~~~(i = 1, 2).     
\label{signa} 
\eea 
 
Let us consider the case  $x > 0$ and  therefore $y > 0$.  
Since both $A_i$ are negative, see Eq.~(\ref{signa}),   
the equality  in Eq.~(\ref{SPeq1}) can be  satisfied only if  
\be 
\sin\theta_{13} =  \,x,\quad A_1 = 0\,, \quad A_2 = 0 \,. 
\label{solinSM} 
\ee 
Then, according to Eq.~(\ref{aa1}), there are two types of solutions of equation  $A_1 = 0$:  
 
\begin{enumerate} 
\item $\kappa = 0$. In this case, we find from Eq.~\eqref{A2},    
$A_2 = \sin \theta_{13}(\sin \delta \sin 2 \theta_{23} - 1) = 0$ which   
gives $\delta = \pi/2$ and $\theta_{23} = \pi/4$. There are no constraints on $\theta_{12}$. 
\item $\theta_{12} = 0$. Then, from Eq.~\eqref{A2}, 
we have $A_2 = \sin \theta_{13}[\sin (\delta - \kappa)\sin 2  
\theta_{23} - 1] = 0$ which is satisfied, if $\theta_{23} = \pi/4$ and  
$\delta - \kappa = \pi /2$. Now only one combination of the two phases  
is determined.  
\end{enumerate} 
In particular, we stress that there exists a mixing matrix 
that satisfies the constraints for every value of $\theta_{12}$ which is in this sense undetermined. 
 
Similarly one can consider the case of $x < 0$ which leads to  
$\sin \theta_{13} = - x$ and changes the signs of the phases.  
So, for both signs of $x$ the first class of solutions can be written  in the following way:  
\be 
\sin\theta_{13} =  \pm \,x\,,\quad \theta_{23} =  \frac{\pi}{4} \,,\quad \delta =  
\pm \frac{\pi}{2} \,,\quad \kappa = 0,   
\label{compatiblesol} 
\ee 
with $\theta_{12}$ and the second Majorana phase $\chi$ are undetermined.  
This type of solution can provide a good first approximation to  
the mixing matrix.  
 
On the other hand, the second type is given by  
\be 
\sin\theta_{13} =  \pm \, x\,,\quad \theta_{23} =  
\frac{\pi}{4} \,,\quad \theta_{12} = 0 \,, 
\quad \delta - \kappa = \pm \frac{\pi}{2} \, .  
\label{incomp} 
\ee 
Clearly the vanishing value of $\theta_{12}$ contradicts observation. However,  
since $m_1 = m_2$, small perturbations which produce splitting can also generate large  
1-2 mixing  (see  Section~\ref{sec3new}).  
 
There is simple interpretation of the two cases considered above 
in terms of the PMNS matrix in Eq.~(24) and the neutrino mass matrix in the flavor basis, Eq.~(2). 
If $\kappa = 0$, then $\Gamma_M = diag(1,1, e^{i\chi})$,  so that 
$U_{12}$ and  $\Gamma_M$ commute. After permutation of these matrices in Eq.~(2) 
$U_{12}$ turns out to be attached to the  degenerate mass matrix. 
Consequently, it can be omitted and we obtain 
\be 
U_{PMNS} = U_{23} \Gamma_\delta U_{13} \Gamma_\delta^* \Gamma_M.  
\label{PMNSb} 
\ee 
Obviously, the same result  Eq.~(\ref{PMNSb}) can be obtained for 
the second case when $\theta_{12} = 0$. Thus, the two solutions, 
we have found, correspond to the elimination of $U_{12}$ matrix.

The two types of solution can be represented in the form of immediate  
relations between the mixing parameters and mass ratios.  
In the first case, Eq.~(\ref{compatiblesol}) we have  
\be 
\sin^2 2 \theta_{23} = \pm \sin \delta  = \cos \kappa = \frac{m_1}{m_2}, 
\label{relA} 
\ee 
and in the second one  
\be 
\sin^2 2 \theta_{23} =  \cos 2 \theta_{12} =  \pm \sin (\delta - \kappa) = \frac{m_1}{m_2}. \label{relB} 
\ee 
In both cases  $\theta_{13}$ is not related to masses. 
Although the relations Eq.~\eqref{relA} and Eq.~\eqref{relB} 
are obtained {\it post factum} their appearance is not accidental: 
symmetry which led to certain values of mixing angles 
and phases encodes information about masses (mass spectrum).

It is not hard to understand why $\theta_{12}$ should be undetermined in the solution of type 1.   
Indeed, for partially  
degenerate spectrum  we have additional freedom to perform an arbitrary rotation  
in the 1-2 plane, $O_{12} = O_{12}(\theta)$.   
In this case the mixing matrix has general form  
$$ 
U_{PMNS} = U_{23}\Gamma_\delta U_{13} \Gamma_\delta^* U_{12} \Gamma_M O_{12}.  
$$ 
If $\kappa = 0$, as is the case in the type 1 solution, then  $\Gamma_M = \trm{diag}\{1, 1, e^{i\chi}\}$, so that   
$\Gamma_M O_{12} = O_{12} \Gamma_M$. Consequently,   
the additional 1-2 rotation is reduced to  
$$ 
U_{12}(\theta_{12}) \rightarrow U_{12}(\theta_{12} + \theta),   
$$ 
where $\theta$ is arbitrary and therefore the 1-2 mixing  
is undetermined. On the other hand, for the type 2 solution  
$\kappa$ can be different from zero. Thus, the freedom to redefine  
$\theta_{12}$ no longer exists and it is natural that a precise  
value for $\theta_{12}$ is predicted, $\theta_{12} = 0$ in this case. 
 
The solutions we obtained correspond to zero values of the left and 
right handed parts of Eqs.~\eqref{eq:mainre} and \eqref{eq:mainim} separately.    
They can be written in a parameterization independent form as 
\begin{align} 
  |U_{\alpha 3}| & = \pm \,x\,, \label{cond1}  \\ 
{\rm Im} \left[U_{\beta 1} U_{\beta 2}^* -  
U_{\gamma 1} U_{\gamma 2}^* \right] & =  \pm \,x \,, \label{cond2} \\ 
\trm{Im}[U_{\alpha 1}U^*_{\alpha 2}] & = 0 \, , \label{cond3} \\ 
 |U_{\beta 3}|^2 & = |U_{\gamma 3}|^2 \, , 
 \label{cond4} 
\end{align} 
being valid  for any choice of flavor index $\alpha$.  
The relation  Eq.~\eqref{cond4} leaves only $\alpha = e$ for a plausible  
explanation of the experimental data.

Let us compute the group parameter $x$ which determines $\theta_{13}$ 
(see Eq.~\eqref{compatiblesol}).    
The combinations  of numbers $(m,\,n) = (3,\,3)$, $(3,\,4)$  
$(4,\,3)$, $(3,\,5)$ and $(5,\,3)$,   
which determine the angles $\psi = 2\pi k/m$ and $\zeta = 
2\pi l /n$,  exhaust the finite von Dyck groups.  We find that the experimental value of $\theta_{13}$ is best  
approximated by choosing $ (m,\,n) = (5,\,3)$ or (3, 5) which  corresponds to the group  
$\mbf{A}_5$. In this case, for $k=2$, $l=1$ we obtain 
\be 
\sin\theta_{13} = \cot\frac{\pi}{3}\cot\frac{2\pi}{5} =  
\sqrt{\frac{1}{3}\left(1-\frac{2}{\sqrt{5}}\right)} \simeq 0.187 
\label{A513}.  
\ee 
This value, $\theta_{13} \approx 11^{\circ}$,  
is a good first approximation to the measured one \cite{globalfit}.  
 
As another example we take $(m,\,n) = (3,\,3)$ with $k=1$ and $l=1$. This  
corresponds to an $\mbf{A}_4$ group and leads to  
\be 
\sin\theta_{13} = \cot^2\frac{\pi}{3} = \frac{1}{3} \label{A413} \, 
\ee 
which has interesting  theoretical implications for the case of complete degeneracy, 
as we see in Sec.~\ref{sec4},  despite being rather far from the experimental value.

\section{Corrections to symmetry results} 
\label{sec3new} 

The mixing and mass splitting we have obtained in the previous sections 
as consequences of symmetry do not agree with experimental data. In particular, 
the 1-2 mass splitting is zero, 
the 1-2 mixing is zero or undefined and the 2-3 mixing is maximal which 
is somewhat disfavored by present data. 
In what follows we will show that actually, 
the obtained mass and mixing patterns can be considered 
as zero order structures. For this we prove that relatively 
small corrections to {\it the neutrino mass matrix} (and not to the mixing) 
can fix the problems listed above. 
 
For definiteness we will consider the effect of a perturbation on the first solution,  
Eqs.~\eqref{compatiblesol}, \eqref{PMNSb} - the second solution can be considered similarly. 
In this case the mixing matrix is given by
\be 
U_{PMNS}^0 = U_{23}(45^{\circ}) \Gamma_{\pi/2} U_{13} \Gamma_{-\pi/2} 
= 
\frac{1}{\sqrt{2}} 
\left( 
\begin{array}{ccc} 
\sqrt{2} c_{13}  &  0  &  - i \sqrt{2} s_{13}  \\ 
- i s_{13}  &   1   & c_{13}  \\ 
- i s_{13}  &   -1  & c_{13} 
\end{array} 
\right). 
\label{pmns0} 
\ee 

In what follows, it will be convenient to consider 
the Majorana phase attached to the third mass eigenvalue. 
That is, we start with a zeroth order mass matrix:
\be
m_{\nu d}^0 \equiv \trm{diag}\{m,\, m,\, m'_\chi\}  
\label{mnuBchi}
\ee
with
$$ 
m^{\prime}_{\chi} \equiv m^{\prime} e^{-i2\chi}. 
$$ 
Let us introduce a perturbation of Eq.~\eqref{mnuBchi}: 
$$ 
m_\nu = m_{\nu d}^0 + \delta m_{\nu d} \,, 
$$ 
where $\delta m_{\nu d}$ is assumed to take a simple form 
$$
\delta m_{\nu d} \equiv {\rm diag} (0, \epsilon, 0). 
$$ 
This perturbation yields the 1-2 mass splitting 
\be 
\Delta m_{21}^2 = 2m\epsilon + \epsilon^2 \epsilon 
\label{epsilon} 
\ee 
and makes the 1-2 rotation physical. Using Eqs.~\eqref{numass} 
and \eqref{pmns0} we obtain in the flavor basis: 
\be 
m_\nu = m^0_\nu + \delta m_\nu, 
\label{eq:summ} 
\ee 
where the zeroth order matrix $m^0_\nu$ 
\be 
m^0_\nu = U_{PMNS}^{0 *} m_d U_{PMNS}^{0 \dagger} = 
m \left( 
\begin{array}{ccc} 
1  - s_{13}^2(1 + r) &  i \frac{1}{\sqrt{2}} s_{13} c_{13}(1 + r) 
&   i \frac{1}{\sqrt{2}} s_{13}c_{13}(1 + r) \\ 
...  &   \frac{c_{13}^2}{2} (1 + r)  &  \frac{c_{13}^2}{2}(1 + r) - 1 \\ 
...  & ... & \frac{c_{13}^2}{2} (1 + r) 
\end{array} 
\right), 
\label{eq:zerom} 
\ee 
and 
\be 
r \equiv \frac{m^{\prime}}{m} e^{-i2\chi}. 
\ee 

The  matrix in Eq.~(\ref{eq:zerom}) has the features that correspond 
to maximal $\theta_{23}$ and vanishing $\theta_{12}$, i.e.,  
$m_{e \mu}^0 = m_{e \tau}^0$ and $m_{\mu \mu}^0 = m_{\tau \tau}^0$. 
The partial degeneracy is encoded in a more complicated relation between 
the elements: $m_{\mu \mu}^0(m_{e e}^0 - m_{\mu \mu}^0 +  m_{\mu \tau}^0) =  m_{e \tau}^{0~ 2}$.  
Violation of these equalities leads to generation of the 1-2 mixing and splitting 
as well as  deviation of the 2-3 mixing from maximal. 
 
The matrix of corrections $\delta m_\nu$ in Eq.~(\ref{eq:summ}) can be written as 
\be 
\delta m_\nu =  \epsilon V \times V^T, 
\label{eqvv} 
\ee 
where $V$ is the second column of the PMNS matrix: 
$$ 
V^T \equiv \left\{ 
s_{12} c_{13},~~ \frac{1}{\sqrt{2}}(c_{12} + i s_{12} s_{13}), 
~~ \frac{1}{\sqrt{2}}(-c_{12} + i s_{12} s_{13} ) \right\}. 
$$ 
Here we left all the parameters unchanged except for the introduction of nonzero 1-2 mixing. 
From Eq.~(\ref{epsilon}) we obtain 
\be 
\epsilon =  \sqrt{\Delta m_{21}^2 + m^2} - m. 
\label{epsilon1} 
\ee 
In the case of strong normal mass hierarchy 
$m \ll \epsilon$ and $\epsilon =  \sqrt{\Delta m_{21}^2}$. 
On the other hand, for strongly degenerate spectrum we have $\epsilon =  \Delta m_{21}^2/2m$. 
The latter expression is also obtained in the case of strong 
inverted mass hierarchy when $m \approx  \sqrt{\Delta m_{31}^2}$.  In this case 
$$ 
\frac{\epsilon}{m} \approx \frac{\Delta m_{21}^2}{2\Delta m_{31}^2} 
\approx 1.6 \cdot 10^{-2}. 
$$ 
 
Comparing the zeroth order values of the elements of the mass matrix, 
Eq.~(\ref{eq:zerom}), with the corrections in Eq.~(\ref{eqvv}) 
we arrive at the following conclusions:

\begin{enumerate}

\item For the $ee-$element, $\delta m_{ee}/m_{ee}^0 \approx 2$ for the 
 normal mass hierarchy when $m \ll \epsilon$.  The ratio goes below 0.4 when 
$m^2 \geq \Delta m_{21}^2$. 
 
\item For the off-diagonal elements in the case of normal mass hierarchy 
we obtain $\delta m_{e\mu}/m_{e \mu}^0 \sim$  
$1/{2 s_{13}}$, $\sqrt{\Delta m_{21}^2/\Delta m_{31}^2} \sim 0.4$ 
for  $m \ll \epsilon$. If $m^2 \geq \Delta m_{21}^2$, the ratio is less than 0.15. 
 
\item In the case of inverted mass hierarchy for $r \ll 1$  the corrections equal  
$\delta m_{ee}/m_{ee}^0 \sim \epsilon/3m \approx 5 \cdot 10^{-3}$ and 
$\delta m_{e\mu}/m_{e\mu}^0 = 
\epsilon/(m \sin 2 \theta_{13}) \approx 5 \cdot 10^{-2}$. 
The corrections for the elements of the $\mu-\tau$ block are of the order $\epsilon/m$. 
\end{enumerate}

Thus, except for the $ee-$elements in  the case of strong 
normal mass hierarchy the relative corrections to the mass matrix required 
to generate 1-2 mass splitting and 1-2 mixing are small:  less than 0.2. 
At the same time, other parameters - the masses, 1-2 and 1-3 mixing 
and the CP-phase - can remain unchanged. The latter however implies 
correlations among the corrections to different elements of mass matrix which might be difficult to achieve. 
 
If generic corrections of order $\epsilon \sim 0.2 m$ are introduced in the mass matrix, 
all the mass and mixing parameters will be modified.
Let us prove that these modifications can be small.
For this we will take the simple perturbation matrix 
\be 
\delta m_{dem} = \frac{\epsilon}{3} \Gamma_\pi D \Gamma_\pi = 
\frac{\epsilon}{3}  \left( 
\begin{array}{ccc} 
1 & 1  &  -1   \\ 
...  & 1  &  - 1  \\ 
...  & ... &  1 
\end{array} 
\right), 
\ee 
where $D$ is the democratic matrix with all elements being 1 and 
$\Gamma_\pi =  {\rm diag}(1, 1, -1)$\footnote{This matrix is close 
to the correction matrix in Eq.~(\ref{eqvv}) and  can be motivated by symmetry arguments.}. 
 
Let us compute the masses and mixing parameters for 
$m'_\nu = m^0_\nu + \delta m_{dem}$, 
where $m^0_\nu$ is given in Eq.~(\ref{eq:zerom}). For the mass 
eigenvalues $m'_i$ of $m'_\nu$, and neglecting contributions of order $s_{13}^2$, we obtain:
\be
m'_1 = m \,,\quad m_2' = m + \frac{1}{3}(2 + c_{13}^2) \simeq m + \epsilon\,,\quad m'_3 \simeq m' \,.
\ee
In order to find the corresponding mixing angles, 
we first make the zeroth order rotation in Eq.~(\ref{PMNSb}) which yields 
\be 
U_{PMNS}^{0 T}  (m^0_\nu + \delta m_{dem}) U_{PMNS}^0 = 
\frac{\epsilon}{3} \left( 
\begin{array}{ccc} 
c_{13}^2 + 3m/\epsilon    &  \sqrt{2} c_{13} & - i s_{13}c_{13} \\ 
...  &   2 +  3m/\epsilon  &  - i \sqrt{2}s_{13} \\ 
...  & ... &  - s_{13}^2 + 3m r/\epsilon 
\end{array} 
\right). 
\label{deltam} 
\ee 
The matrix  above can be subsequently diagonalized by a rotation
\be 
U^{\prime} = \Gamma_{\pi/2}U_{13}^{\prime} U_{23}^{\prime} U_{12}^{\prime} \Gamma_{-\pi/2}.  
\label{uprime}
\ee
Up to order $s_{13}^2$ corrections and other small angles corrections, 
this  gives $\sin^2 \theta_{12}^{\prime} \approx 1/3$ 
in good agreement with data. Furthermore, 
if we assume for simplicity that $\chi = 0$ so that $r$ is real, 
and multiply Eq.~(\ref{deltam}) by $\Gamma_{\pi/2}$ which follows from Eq.~(\ref{uprime})  
we obtain  
$$ 
\tan \theta_{13}^{\prime} \approx  - \frac{\epsilon}{3 m^{\prime}} s_{13} c_{13} 
\leq s_{13} \frac{1}{3}\sqrt{\frac{\Delta m_{21}^2}{2\Delta m_{31}^2}} \sim 0.05 s_{13}, 
$$ 
and 
$$
\tan \theta_{23}^{\prime} \approx - \frac{1}{\sqrt{2}} \frac{2s_{13}}{3}\frac{\epsilon}{ m^{\prime}} 
\sim \frac{0.02}{\sqrt{2}}, 
$$ 
{\it i.e.} less than $2\%$. 
Thus,  the PMNS matrix including corrections can  be written as 
\bea 
U_{PMNS} = U_{PMNS}^0 U^{\prime} =  
U_{23}(45^{\circ}) \Gamma_{\pi/2} U_{13} (\theta_{13} + \theta_{13}^{\prime}) 
U_{23}^{\prime} U_{12}^{\prime} \Gamma_{-\pi/2} 
\nonumber\\ 
= U_{PMNS}^0 (\theta_{13} + \theta_{13}^{\prime}) 
\Gamma_{\pi/2} U_{23}^{\prime} U_{12}^{\prime} \Gamma_{-\pi/2} . 
\label{mixcorr} 
\eea 
The PMNS matrix is determined by Eq.~(\ref{mixcorr}) up to a phase 
matrix which can be attached from the right and we will use this to reduce the 
 Eq.~(\ref{mixcorr}) to standard parametrization form. 
We can now compute the elements of the matrix in Eq.~(\ref{mixcorr}) explicitly 
and identify them with the elements of the mixing matrix in the standard 
parameterization (subscripts $s$). 
The $e2-$element equals $c_{13}^s s_{12}^s = 
c_{13} s_{12}^{\prime} - s_{13} c_{12} s_{23}^{\prime}$, that is,  
the correction to the equality $\theta_{12}^s = \theta_{12}^\prime$ is of the order 
$s_{13}s_{23}^{\prime}$. 
In order to determine other angles and the Dirac CP phase it is enough to consider the third  
column of Eq.~(\ref{mixcorr}): 
\be 
(U_{PMNS})_{\alpha 3}^T = 
\left(- i s_{13}c_{23}^{\prime} e^{-i \pi/2 },~ \rho e^{i \xi},~ \rho e^{- i\xi} 
\right), 
\label{3col} 
\ee 
where 
\be 
\rho = \frac{1}{\sqrt{2}} \sqrt{c_{13}^2 c_{23}^{\prime 2} + s_{23}^{\prime 2}} 
\approx \frac{1}{\sqrt{2}}, 
~~~ \tan \xi = - \frac{1}{c_{13}} \tan \theta_{23}^{\prime} 
\ee 
or $\xi \approx - \theta_{23}^{\prime}$. The phase of the  $U_{\mu3}-$element in Eq.~(\ref{3col}) 
can be removed {(as it should be in the standard parametrization)} 
by acting on the right hand side of Eq.~(\ref{mixcorr}) 
{\bf with}  the additional phase matrix ${\rm diag}(1, 1, e^{-i \xi})$. This means that  the CP-phase 
is modified to $\delta = \pi/2 + \xi$. 
 
Thus, we have shown that a simple correction matrix can generate an acceptable 1-2 mixing, the required mass splitting and produces only small (few per cent) corrections to the other mixings and to the CP-violation phase.

 
\section{Constraints on mixing for the completely degenerate spectrum} 
\label{sec4} 
 
 
As we remarked in Sec.~\ref{sec2}, 
$M_{\nu U}$ can be forced to be completely degenerate, if a non-abelian discrete  
subgroup of $O(3)$ with three-dimensional representations  
is imposed as $G_\nu$. The possible groups are thus restricted  
to $\mbf{A}_4$, $\mbf{S}_4$ and $\mbf{A}_5$.  
These can be generated by two matrices:  
$S_\zeta$ and $P$ that satisfy 
\be 
S_\zeta^n = P^2 = (S_\zeta P)^r = \mbb{I} \,. 
\ee 
We take a basis for the  neutrinos such that $S_\zeta$ is given by  
Eq.~\eqref{SB}. The second matrix, $P$,  can be  represented as  
\be 
P = O^TP_DO ,  
\ee 
where 
\be 
P_D = \trm{diag}\{1,\,-1,\,-1\} \,,  
\ee 
and $O = O(\phi_{12}, \phi_{13}, \phi_{23})$ is a generic orthogonal matrix  
of rotations on the angles $\phi_{ij}$.  
 
In the charged lepton sector we take, as before,   
$G_\ell = \mbf{Z}_m$.  
The generator $T$ must now satisfy conditions like Eq.~\eqref{wrelation} with both $S_U$  
and $P_U = U_{PMNS}PU^\dagger_{PMNS}$. Hence, the complete presentation for the  
flavor group $G_f$ is given by 
\begin{align} 
S_U^n & = T^m  = P^2_U  = \mbb{I}\,,\quad (S_UP_U)^r =  
(S_\zeta P)^r = \mbb{I}\,, \label{pr1} \\ 
& \quad \quad \quad (S_UT)^{2} = (P_UT)^q = \mbb{I} \label{pr2} \,. 
\end{align} 
Notice that this presentation does not guarantee that $G_f$ is finite.  
Following the same argument exploited  
in case B, we obtain that Eqs.~(\ref{pr1}, \ref{pr2}) impose a set of conditions  
on matrices $U_{PMNS}$ and $O$:  
\begin{align} 
\tr{U_{PMNS}S_\zeta U^\dagger_{PMNS}T} & = - 1, 
\label{C1} \\ 
\tr{OS_\zeta O^T P_D} & = a_r,   
\label{C2} \\ 
\tr{U_{PMNS}O^T P_D O U^\dagger_{PMNS}T} & = a_q,   
\label{C3} 
\end{align} 
where $a_r$($a_q$) is the sum of three $r$-th ($q$-th) roots of unity.  
The solutions of  Eq.~\eqref{C1}, which coincides with condition of the  
previous case, are given in Eqs.~\eqref{compatiblesol}, \eqref{incomp}. 
Two other equations are new:  
Eq.~\eqref{C2} is  the one for the matrix $O$, instead of $U_{PMNS}$, and  
it can be solved in a  
similar way. Using a parametrization  
for matrix $O$ similar to Eq.~\eqref{pmnspar} with vanishing CP phases  
we get 
\be 
\sin^2\phi_{13} = \frac{a_r + 1}{2(1+\cos\zeta)} \,,  
\label{phi13} 
\ee 
where $\phi_{13}$ is the angle in $O$ equivalent to $\theta_{13}$ in Eq.~\eqref{pmnspar}.  
Substituting  \eqref{compatiblesol}  
and \eqref{phi13} in Eq.~\eqref{C3}, we obtain a new equation  
for the remaining parameters of $U_{PMNS}$ that either  
has no solution - and the group representation  in question does not exist - or fixes  
the Majorana phase $\chi$.

For the values of the parameters in Eqs.~\eqref{compatiblesol} 
and \eqref{A513} the  Eq.~\eqref{C3} has no   
solutions. For  the  
pattern with the 1-3 mixing from  Eq. (\ref{A413}) 
the  Eq.~\eqref{C3}  does have a solution if  $r=3$ and  
$q=3$. that for $r = 3$ the group   $G_\nu = \mbf{A}_4$. We obtain for  
the second Majorana phase, $\chi$: 
\be 
\chi = \frac{3\pi}{2} \label{chiA4a}.  
\ee

A few comments are in order.  
It is easy to check that $T$ can be written as  
a combination of $P_U$, and $S_U$,  so that it is not an  
independent generator. Since $G_f = \mbf{A}_4$,  
this theory corresponds to a case in which the  
flavor group $G_f$ remains \emph{unbroken}  
in the neutrino sector while it is broken to a $\mbf{Z}_3$  
subgroup in the charged lepton sector.

Out of six parameters  
that appear in $U_{PMNS}$ three are unphysical in the fully  
degenerate case~\cite{degeneratenus}. This seems to be in contradiction with the fact  
that we have determined five parameters  
$\{ \theta_{13},\, \theta_{23},\,\delta,\,\kappa,\,\chi \}$ by means of the  
symmetry.  
Actually, some of these parameters have been fixed,  
not by the symmetry but by our choice of basis.  
Indeed, in order to perform the analysis, we assumed  
that the group $G_f = \mbf{A}_4$ included the 1-2 rotation $S_\zeta$  
as one of the generators. However, for fully degenerate neutrinos,  
rotations around any axis could serve as symmetries of the neutrino mass matrix.  
Hence, if $S_U$,  $P_U$ and $T$ satisfy Eqs.~\eqref{C1} and \eqref{C2}  
for some $U_{PMNS}$, then also $T$ and the new matrices $S'_U$, $P'_U$   
defined as 
\be 
S'_U = VS_UV^T \,,\quad P'_U = VP_UV^T 
\ee 
satisfy Eqs.~\eqref{C1} and \eqref{C2} for a mixing matrix $U'_{PMNS}$ given by  
\be 
U'_{PMNS} = U_{PMNS}V^T \label{newU}.   
\ee 
Here $V$ is any orthogonal matrix.  Thus, the mixing parameters found are written in basis-dependent form. One can only say that there exists a 
basis in which the $U_{PMNS}$ angles and phases have the values in  
Eqs.~\eqref{compatiblesol} and \eqref{chiA4a}.  In general, according to  
Eq.~\eqref{newU}, three parameters are unphysical out of the six that appear in $U_{PMNS}$ in the standard parametrization.  
 
The basis-independent physical quantities are combinations of the elements of $U_{PMNS}$ that are invariant under orthogonal rotations of the neutrino fields and the usual phase redefinitions of leptons. These functions are nothing else but the absolute values of the elements of the matrix $\mcl{U} = U_{PMNS}^T U_{PMNS}$. It is easy to see that since $\mcl{U}$ is symmetric and unitary, only 3 out of the 9 elements $|\mcl{U}_{ij}|$ are independent as expected according to the analysis above. Furthermore, the matrix $\mcl{U}$ is proportional to the mass matrix in the flavor basis which has physical meaning, e.g. its $ee$-element determines the amplitude on neutrinoless double-beta decay. 

 
\section{Conclusions}  
\label{sec5} 
 
 
\vth 
In this paper, we further developed the formalism  
of the ``symmetry building'' in such a way that it includes  
both mixing parameters and  neutrino masses. 
More precisely, the formalism  connects  
partially and completely degenerate neutrino spectra  with  the mixing angles  and CP-phases.  
These are the only possibilities (along with zero mass) which can be obtained as consequences of the  
unitary residual symmetries.

The case of partial degeneracy,  $m_1 = m_2$,  
follows when a $\mbf{Z}_n$ subgroup of $SO(2)$ with $n\geq3$ is preserved in the neutrino sector. It 
can be a good lowest order approximation to the spectrum of normal 
(inverted) mass hierarchy. This case is very  
restrictive, leading to 4 conditions on 
the mixing parameters. For $m_1 = m_2$  
we have found two types of solutions with 4 mixing parameters  
fixed. Both solutions show maximal 2-3 mixing and  
1-3 mixing determined directly by   
the group parameters. They  differ by  
the values of the 1-2 mixing and CP-violation phases.   
The first solution has zero $\theta_{12}$, and one condition on the phases:  
$\delta - \kappa =  \pi/2$. In the second solution, $\theta_{12}$ is undefined  
but both phases are fixed: $\delta = \pi/2$ and $\kappa = 0$.  
In the case that gives the best approximation  
to the measured values, the symmetry group is $\mbf{A}_5$ and we obtain $\sin \theta_{13} = 0.187$.  
 
These solutions should be considered as a lowest order approximation. 
Relatively small corrections can  produce the mass splitting and  
fix $\theta_{12}$  in one case and generate $\theta_{12}$ in another.   
Corrections may also give better agreement of the 1-3 and 2-3 mixings 
with observations.  We show that in the first case the corrections proportional to  the  
``democratic'' matrix can produce the 1-2 mass splitting  
and mixing in agreement with observations  
while giving rise to very small corrections to the other mixing  parameters and CP-phases.
 
A completely degenerate spectrum is achieved if the residual  
symmetry in the neutrino sector is either $\mbf{A}_4$, $\mbf{S}_4$ or $\mbf{A}_5$.  
In this case, $U_{PMNS}$ has only 3 physical parameters all  
of which are determined by the symmetry.  
In our formalism, this is made explicit by the fact that,  
in a particular basis, all the angles and CP-phases of the mixing matrix  
- except for $\theta_{12}$ which remains undefined -  are fixed. 
 
The values of the charged lepton masses are not involved in this consideration. 
In fact, the inclusion of charged leptons may  produce 
corrections which will make the scheme with degeneracy to be viable. 
At the same time, it will be probably difficult to immediately extend this consideration to  
the quark sector and treat two light families as being degenerate in the first  
approximation.

 
\section*{Acknowledgements} 
 
 
A. Y. S. is grateful to the MPI fur Kernphysik, Heidelberg,   
where this work has been accomplished,  for hospitality. We acknowledge partial support by European Union FP7 ITN INVISIBLES (Marie Curie Actions, PITN-GA-2011-289442)


\end{document}